\documentclass[10pt,twocolumn]{article}
\def\thebibliography#1{\section*{ }
 \list{[\arabic{enumi}]}{\settowidth\labelwidth{[#1]}\leftmargin\labelwidth
 \advance\leftmargin\labelsep\usecounter{enumi}}
 \def\newblock{\hskip .11em plus .33em minus -.07em}\sloppy
 \sfcode`\.1000\relax}

\begin{document}

\title{\large\bf Elementary Particles and Spin Representations}

\author{\small Paolo Maraner\\[-1mm]
       {\small\em Liceo Classico ``G.Carducci", via Manci 8}\\[-1mm]
       {\small\em I-39100 Bolzano, Italy}\\[-1mm]
       {\footnotesize paolo.maraner@comm2000.it}\\[3mm]
\begin{minipage}{14cm}
\footnotesize \noindent We emphasize that the group-theoretical
considerations leading to $SO(10)$ unification of electro-weak and
strong matter field components naturally extend to space-time
components, providing a truly unified description of all
generation degrees of freedoms in terms of a single chiral spin
representation of one of the groups $SO(13,1)$, $SO(9,5)$,
$SO(7,7)$ or $SO(3,11)$. The realization of these groups as higher
dimensional space-time symmetries produces unification of all
fundamental fermions is a
single space-time spinor.\\[1mm]
\noindent {\em Keywords:} elementary particles; fermion
unification; extra dimensions. \\[1mm]
\noindent {\em PACS numbers:} 12.10.-g; 11.30.Ly; 04.50.+h
\end{minipage}}
\date{}
\maketitle

\noindent Nowadays, it is largely accepted that at some
intermediate energy scale elementary matter and non-gravitational
fundamental interactions should go through an $SO(10)$ grand
unification \cite{Georgi74,essentialSo10}. While it is open the
question on what the detailed dynamics at that scale should be, we
are very confident that the group theoretical framework is
correct. Our belief mainly grounds on the fact that the
complicated family structure reorganizes in a single (spin)
representation. This simplicity is a primary attractive of the
model. Unfortunately, $SO(10)$ only offers a partial answer to the
unification quest: {\sl 1.} the family representation has to be
supported by its complex conjugate to take into account
antiparticles; {\sl 2.} the couple of family/anti-family
representations has to be replicated |at least three times| to
take into account generations; {\sl 3.} last but not the least,
elementary fermions transform as composite objects under
space-time and internal gauge transformations, a signal that
gravity has not been taken into account. Over the years, many
different ideas have been proposed to overcome this situation
\cite{beyondSo10}. This paper presents an original viewpoint. We
emphasize that the group-theoretical considerations leading to
unification of electro-weak and strong matter filed components
naturally extend to space-time components, providing a truly
unified description of all fundamental fermions in terms of a
single chiral spin representation of a fourteen dimensional
pseudo-orthogonal space-time group. Our consideration are mainly
of cinematical nature, only providing a general framework in which
the whole unification issue may possibly be reconsidered. We will
only touch on the dynamical problem. In order to better illustrate
our viewpoint, we start by focusing on the family structure and
resume the key observations that lead to treat leptons and quarks
on a similar footing.
\\
Pati-Salam \cite{Pati-Salam74} idea of lepton number as a fourth
color is briefly summarized by recalling that the chiral spin
representations\footnote{Throughout this paper we use the common
notation of labelling representations by their dimension (in
boldface). Negative chirality Weyl representations are
distinguished by a prime ${\mathbf '}$ or a bar $\bar{\mathbf\ }$
when conjugate to the positive chirality ones.} ${\mathbf 2}$ and
${\mathbf 2}'$ of $SO(4)\simeq SU_L(2)\times SU_R(2)$ respectively
transform as ${\mathbf 1}+{\mathbf 1}$ and ${\mathbf 2}$ under the
subgroup $SU_L(2)$, while the chiral spin representations
${\mathbf 4}$ and $\bar{\mathbf 4}$ of $SO(6)\simeq SU(4)$
respectively transform as ${\mathbf 1}+\bar{\mathbf 3}$ and
${\mathbf 1}+{\mathbf 3}$ under the regularly embedded $SU(3)$. By
inspecting product representations
\begin{equation}
\begin{array}{l}
({\mathbf 2},{\mathbf 4})
=({\mathbf 1},{\mathbf 1}) + ({\mathbf 1},{\mathbf 1}) +
({\mathbf 1},\bar{\mathbf 3})+({\mathbf 1},\bar{\mathbf 3})\\
({\mathbf 2},\bar{\mathbf 4})
=({\mathbf 2},{\mathbf 1})+({\mathbf 2},{\mathbf 3})\\[1mm]
({\mathbf 2}',{\mathbf 4})
=({\mathbf 1},{\mathbf 1}) + ({\mathbf 1},{\mathbf 1}) +
({\mathbf 1},{\mathbf 3})+({\mathbf 1},{\mathbf 3})\\
({\mathbf 2}',\bar{\mathbf 4})
=({\mathbf 2},{\mathbf 1})+({\mathbf 2},\bar{\mathbf 3})
\end{array}
\end{equation}
we realize that the complicated $U(1)\times SU(2)\times SU(3)$
Standard Model matter representation appears much simpler in terms
of $SO(4)\times SO(6)$ (hypercharge also can be accommodated). The
enlargement of the gauge group |without addition of matter
freedoms| produces a remarkable simplification in the theory's
structure, allowing to treat leptons and quarks on the very same
ground. Yet, the situation is not completely  satisfactory. The
elementary matter representation still is composite and the gauge
group still has a direct product structure. The problem was
overcome by Georgi \cite{Georgi74}, essentially by recalling that
the chiral spin representations of the ten dimensional orthogonal
group $SO(10)$ |also the ones of the pseudo-orthogonal group
$SO(4,6)$ do| transform as
\begin{equation}
\begin{array}{l}
{\mathbf{16}}= ({\mathbf 2},{\mathbf 4})+({\mathbf 2}',\bar{\mathbf 4})\\
\bar{\mathbf{16}} =({\mathbf 2},\bar{\mathbf 4})+({\mathbf 2}',{\mathbf 4})
\end{array}
\end{equation}
under the regular subgroup $SO(4)\times SO(6)$. The further
enlargement of the gauge group, still without addition of matter
freedoms, produces unification inside each family. All left-handed
matter fields ${\nu_R}^c$, $(\nu,e)_L$, ${e_R}^c$, ${u_R}^c$,
$(u,d)_L$ and ${d_R}^c$ appear as  components of the single
irreducible representation ${\mathbf{16}}$ while all right-handed
matter $\nu_R$, ${(\nu,e)_L}^c$, $e_R$, $u_R$, ${(u,d)_L}^c$ and
$d_R$ makes up its conjugate $\bar{\mathbf{16}}$. The gauge group
is no longer a direct product. As far as electro-weak and strong
interactions are concerned and regardless to the problems
mentioned above, $SO(10)$ grand unification appears so natural to
seem unavoidable.

Besides indices labelling particles properties under internal
gauge transformations, elementary matter fields carry space-time
indices. In the framework discussed above, left-handed matter
transforms according the chiral spin representation $\bar{\mathbf
2}$ of the space-time Lorentz group $SO(3,1)$ and |independently|
according the chiral spin representation ${\mathbf{16}}$ of the
grand unified gauge group $SO(10)$; right-handed matter transforms
according ${\mathbf 2}$ under space-time transformations and
according $\bar{\mathbf{16}}$ under internal gauge
transformations. In this prospective the theory does not look much
unified. Rather, its structure resembles the one of the
$SO(4)\times SO(6)$ model before $SO(10)$ grand unification: the
elementary matter representation is composite and the group of
allowed transformations present the direct product structure
$SO(3,1)\times SO(10)$.
A possible objection is that, unlike in the
Pati-Salam model, the two groups play different roles. The gauge
group only acts on matter filed components
$\delta\psi={1\over2}\epsilon_{ij}\Sigma^{ij}\psi$ not involving
coordinate transformations. The space-time group involves both,
field components and coordinates transformations. In the special
relativistic context, in which GUTs are in general considered, the
theory is assumed to be invariant under the combined Lorentz
transformation $\delta x^{\mu}=\epsilon^\mu_{\
\nu}x^\nu+\tau^\mu$,
$\delta\psi={1\over2}\epsilon_{\mu\nu}\Sigma^{\mu\nu}\psi$ with
identical infinitesimal parameters $\epsilon^{\mu\nu}$. Taking
into account gravity is equivalent
to gauge the Lorentz group \cite{Kibble61} by considering the
generalized transformations in which the parameters
$\epsilon^{\mu\nu}$ ad $\tau^\mu$ become arbitrary functions of
the coordinates. It is then possible to regard as independent
$\epsilon^{\mu\nu}$ and $\xi^\mu\equiv \epsilon^\mu_{\
\nu}x^\nu+\tau^\mu$. In this way, we can consider generalized
transformations with $\xi^\mu=0$ but arbitrary
$\epsilon^{\mu\nu}$. In the general relativistic context,
coordinates and Lorentz field transformations become completely
independent. The action of SO(3,1) on matter field components
is absolutely analogue to the one of an internal gauge group. \\
It looks then natural to
pursue a unified description of elementary matter
by recalling that the left-handed chiral spin representation of
the fourteen dimensional pseudo-orthogonal group transforms as
\begin{equation}
{\mathbf{64}}' =(\bar{\mathbf 2},{\mathbf{16}})+
                ({\mathbf 2},\bar{\mathbf{16}})
\end{equation}
under the regular subgroup $SO(3,1)\times SO(10)$. For the
signature a few different choices are possible. We can add the 10
Euclidean directions to the 3 space-like ones or the to single
time-like one. Moreover, since we have to deal with a non-compact
group anyway, there is no longer reason to prefer $SO(10)$ to
$SO(4,6)$ in unifying $SO(4)\times SO(6)$. In this case we also
can add the 4 to the 3 and the 6 to the 1 or the 4 to the 1 and
the 6 to the 3. In brief, the fourteen dimensional
pseudo-orthogonal groups
\begin{equation}
SO(13,1), \hskip0,2cm SO(9,5), \hskip0,2cm  SO(7,7), \hskip0,2cm
SO(3,11) \label{so14}
\end{equation}
present $SO(3,1)\times SO(4)\times SO(6)$ as regular subgroup and
correctly contain all generation degrees of freedoms |space-time
and gauge, particles and antiparticles| in their
self-(pseudo)conjugate left-handed chiral spin representation. A
generation of elementary particles is described by a single filed
$\psi_{\mathbf{64}'}(x)$ living in ${\mathbf{64}'}$.\footnote{Some
of these degrees of freedom have to be identified with ordinary
spin.} The eventual enlargement of the gauge group produces a
genuine unification inside each generation, still without addition
of matter freedoms.
\\
At least from the group-theoretical viewpoint, these
considerations give a satisfactory solution to the first and third
problem we mention in the introduction. The second problem,
concerning families replication, apparently remains unsolved.
Should we further extend the gauge group to include the missing
degrees of freedoms? In order to answer this question we leave for
a while the problem of elementary matter unification and turn our
attention to fundamental interactions. The gauge groups of
gravitational, hyper, weak and strong forces are regular subgroups
of the fourteen dimensional pseudo-orthogonal groups (\ref{so14})
and their embedding is in some sense minimal. After symmetry
breaking, {four} directions |identified with the physical
space-time| transform under the Lorentz group; the remaining
{four}+{six} |corresponding to an unphysical internal space| are
respectively needed to realize $SU(2)\subset SO(4)$ and
$SU(3)\subset SO(6)$ transformations (hypercharge is related to
the relative rotation of weak and strong directions). From the
point of view of fundamental interactions a further extension of
the gauge group is unnecessary, if not undesirable. On the other
hand, we still have to specify the sense in which the unified
group has to be understood as gauge group. Given the
pseudo-orthogonal structure and the fact that four of its defining
directions form the physical space-time, the only reasonable
choice seems that of accepting the remaining ten directions as
more physical dimensions. We are lead to a fourteen dimensional
space-time \cite{14D} with a pseudo-Euclidean geometry described
by one of the groups (\ref{so14}). To make contact with low energy
physics it is convenient to introduce local coordinates $x^\mu$,
$\mu=0,1,2,3$ and $\xi^i$, $i=1,2,...,10$, respectively
parameterizing directions transforming under the subgroups
$SO(3,1)$ and $SO(4)\times SO(6)$. The former are identified with
ordinary space-time coordinates, the latter parameterize extra
dimensions. Let us now go back to elementary matter and the
families replication problem. At low energies a generation of
elementary particles is described by a filed
$\psi_{\mathbf{64}'}(x)$ only depending on ordinary space-time
coordinates. In the higher dimensional model such a filed simply
is  a left-handed space-time spinor, but |as any other higher
dimensional filed| this filed depends on {\em all} the space-time
coordinates, the ordinary $x$ and the extra $\xi$
\begin{equation}
\Psi_{\mathbf{64}'}(x,\xi)
\label{matter}
\end{equation}
For every given value of extra coordinates
$\psi_{\mathbf{64}',\xi}(x)\equiv\Psi_{\mathbf{64}'}(x,\xi)$
exactly contains the fields describing a generation of elementary
particles. The whole left-handed space-time spinor includes
infinite many copies of the generation structure. The realization
of the unified group as space-time symmetry makes unnecessary any
further enlargement of the gauge group. {\em All fundamental
fermions are placed in a single representation of the space-time
group.}
\\
The extension of space-time offers a non-group-theoretical
explanation of the existence of many copies of the generation
structure, addressing a possible solution of the families
replication problem. The crucial point remains that of
understanding how many of these copies survive when space-time
symmetry is broken and the number of dimensions is effectively
reduced to four. The answer depends on the particular dynamical
mechanism employed in breaking the symmetry and goes beyond the
goal of the this paper. We note however, that the needed symmetry
breaking can occur in the fermion-fermion operator. The product of
two left-handed spinors decomposes as
\begin{equation}
{\mathbf{64}'}\times{\mathbf{64}'}=
\Lambda_1+\Lambda_3+\Lambda_5+\Lambda_7^-
\label{64-64}
\end{equation}
in terms of $k$-forms $\Lambda_k$, the 7-form being
anti-self-dual. The associated curvature forms |respectively the
closed 2-, 4-, 6- and 8-forms constructed by exterior
differentiation| foliate space-time in 2-, 4- and 6-dimensional
hyper-surfaces. This is exactly what we need to break the
space-time group down to an effective $SO(3,1)\times U(2) \times
U(3)$ through a Freund-Rubin \cite{Freund-Rubin80} like
mechanism.\footnote{The 14-dimensional space-times group can be
broken in different ways in direct products of (pseudo)orthogonal
and (pseudo)unitary symmetries; (pseudo)unitary groups are
obtained when the 2-form survives on 4- and 6-hyper-surfaces,
providing a complex structure.} In addition equation (\ref{64-64})
indicates that chiral fourteen-dimensional spinors are massless.
In brief, the 14-dimensional left-handed space-time spinor
(\ref{matter}) might be the right object describing elementary
matter. In addressing a dynamical model different options are
open. A particulary intriguing one is to assume that
$\Psi_{\mathbf{64}'}$ is the only elementary filed, while the
higher dimensional metric structure |hence the low energy graviton
and gauge bosons| arises as a collective excitation of
fermion-fermion pairs\cite{composite-bosons,spin-gravity}.

 We conclude by considering the problem of discerning among the four
possible choices of unified gauge group. A partial answer comes
from the inspection of spin representations properties
\cite{spin}. For signatures (13,1) and (9,5) the spin
representations carry a quaternionic structure. On
${\mathbf{64}'}$ it only is possible to define a
pseudo-conjugation ${\tilde{\cal C}}$, ${\tilde{\cal C}}^2=-1$,
commuting with the action of the group. For signatures (7,7) and
(3,11) the spin representations  carry a real structure instead. A
genuine operation ${\cal C}$ of complex conjugation, ${\cal
C}^2=1$, is defined. It is compatible with the group action and
commutes with the chiral operator. As a consequence the real and
imaginary parts of ${\mathbf{64}'}$ transform independently,
forming irreducible real (Majorana) representations of the group.
Particles and antiparticles live in a superposition of real and
imaginary parts and the ones appear as the complex conjugate of
the others. This is the common situation dealt with in field
theory. The absence of a properly defined complex conjugation
makes $SO(13,1)$ and $SO(9,5)$ not particularly attractive as
unified groups |if not ruling them out. On the other hand, the
spin representations of $SO(7,7)$ and $SO(3,11)$ are algebraically
undistinguishable and a preference can not be expressed merely on
this ground. From a purely aesthetical viewpoint, the total
symmetry between components with opposite signature makes of
$SO(7,7)$ a preferential candidate. In spite of technical
problems, the idea of a space-time with an equal number of space
and time dimensions, therefore treating space and time really on
the same ground, appears particularly attractive.\\[3mm]

\hrule

\end{document}